\documentclass[aps,twocolumn,prb,eqsecnum,showpacs]{revtex4}

\usepackage{graphicx}
\begin{document}
\draft
\preprint{18 June 2007}
\title{Optical characterization of platinum-halide ladder compounds
       \footnote[2]{to be published in
                    Phys. Rev. B {\bf 76}, November 15, No. 19 (2007)}}
\author{Shoji Yamamoto and Jun Ohara}
\address{Department of Physics, Hokkaido University,
         Sapporo 060-0810, Japan}
\date{18 June 2007}
\begin{abstract}
New varieties of quasi-one-dimensional halogen ($X$)-bridged
transition-metal ($M$) complexes,
(C$_8$H$_6$N$_4$)[Pt(C$_2$H$_8$N$_2$)$X$]$_2X$(ClO$_4$)$_3\cdot $H$_2$O
($X=\mbox{Br},\mbox{Cl}$) and
(C$_{10}$H$_8$N$_2$)[Pt(C$_4$H$_{13}$N$_3$)Br]$_2$Br$_4\cdot 2$H$_2$O,
comprising two-leg ladders of mixed-valent platinum ions, are described
in terms of a multiband extended Peierls-Hubbard Hamiltonian.
The polarized optical conductivity spectra are theoretically reproduced
and the ground-state valence distributions are reasonably determined.
The latter variety, whose interchain valence arrangement is out of phase,
is reminiscent of conventional $M\!X$ single-chain compounds, while the
former variety, whose interchain valence arrangement is in phase, reveals
itself as a novel $d$-$p$-$\pi$-hybridized multiband ladder material.
\end{abstract}
% 02.20.$-$a: Group theory
% 42.65.-k: Nonlinear optics
% 42.65.Tg: Optical solitons; nonlinear guided waves
% 63.20.Kr: Phonon-electron and phonon-phonon interactions
% 71.10.Hf: Non-Fermi-liquid ground states, electron phase diagrams and
%           phase transitions in model systems
% 71.23.An: Theories and models; localized states
% 71.35.-y: Excitons and related phenomena 
% 71.35.Aa: Frenkel excitons and self-trapped excitons 
% 71.38.-k: Polarons and electron-phonon interactions
% 71.45.Lr: Charge-density-wave systems
% 71.55.-i: Impurity and defect levels
% 75.30.Fv: Spin-density waves
% 75.40.Mg: Numerical simulation studies
% 78.20.Bh: Theory, models, and numerical simulation
% 78.20.Ci: Optical constants (including refractive index,
%           complex dielectric constant, absorption, reflection and
%           transmission coefficients)
% 78.30.-j: Infrared and Raman spectra
% 78.47.+p: Time-resolved optical spectroscopies and other ultrafast
%           optical measurements in condensed matter 
% 78.55.-m: Photoluminescence, properties and materials 
\maketitle

\section{Introduction}

   Quasi-one-dimensional transition-metal ($M$) complexes with bridging
halogens ($X$) \cite{M5758,M5763,G6408,W6435} have been attracting much
interest for several decades and significant efforts are still devoted to
fabricating their new varieties.
Conventional platinum-halide chains exhibit a Peierls-distorted
mixed-valent ground state, \cite{C475} while their nickel analogs have a
Mott-insulating monovalent regular-chain structure. \cite{T4261,T2341}
Palladium-halide chains are intermediates with a ground state tunable
optically \cite{I241102,M123701} and electrochemically.
\cite{M7699,M035204}
The charge-density-wave (CDW) ground state can be tuned by halogen doping
\cite{M5593,H5706,Y422} and pressure application \cite{K18682} as well.
Metal binucleation leads to a wider variety of electronic states.
\cite{Y125124,K2163}
A diplatinum-iodide chain compound,
 [(C$_2$H$_5$)$_2$NH$_2$]$_4$[Pt$_2$(P$_2$O$_5$H$_2$)$_4$I],
exhibits photo- and pressure-induced phase transitions,
\cite{S1405,Y140102,Y1489,M046401,Y075113} whereas its analog without
any counter ion,
 Pt$_2$(CH$_3$CS$_2$)$_4$I,
is of metallic conduction at room temperature and undergoes successive
phase transitions \cite{K10068,Y1198} with decreasing temperature.
There are further attempts \cite{S8366,Y6596} to bridge polynuclear and/or
heterometallic units by halogens.

   More than three hundred $M\!X$ compounds have thus been synthesized so
far, but their crystal structures are all based on $M\!X$ single chains.
In such circumstances, several authors \cite{K12066,K7372} have recently
succeeded in assembling $M\!X$ complexes within a ladder lattice.
Metal-oxide ladders are generally remarkable for their strongly correlated
$d$ electrons.
SrCu$_2$O$_3$ behaves as a $d$-$p$ ladder of the Hubbard type,
\cite{N245109} whereas NaV$_2$O$_5$ is well describable within a
single-band Holstein-Hubbard Hamiltonian. \cite{A245108}
On the other hand, the newly synthesized metal-halide ladders are
double-featured with competing electron-electron and electron-phonon
interactions \cite{F044717} and are possibly of $d$-$p$-$\pi$-mixed
character.
Such a multicolored stage potentially exhibits a variety of electronic
states and it is highly interesting to control them chemically and
physically.
Thus motivated, we make a model study of ladder-shaped $M\!X$ compounds,
($\mu$-bpym)[Pt(en)$X$]$_2X$(ClO$_4$)$_3\cdot$H$_2$O
($X=$Br,Cl;
 en$=$ethylendiamine$=$C$_2$H$_8$N$_2$;
 $\mu$-bpym$=2,2'$-bipyrimidine$=$C$_8$H$_6$N$_4$)
and
(bpy)[Pt(dien)Br]$_2$Br$_4\cdot 2$H$_2$O
(dien$=$diethylentriamine$=$C$_4$H$_{13}$N$_3$;
 bpy$=4,4'$-bipyridyl$=$C$_{10}$H$_8$N$_2$),
which are hereafter referred to as
(bpym)[Pt(en)$X$]$_2$ and (bpy)[Pt(dien)Br]$_2$, respectively.
\begin{figure}
\centering
\includegraphics[width=85mm]{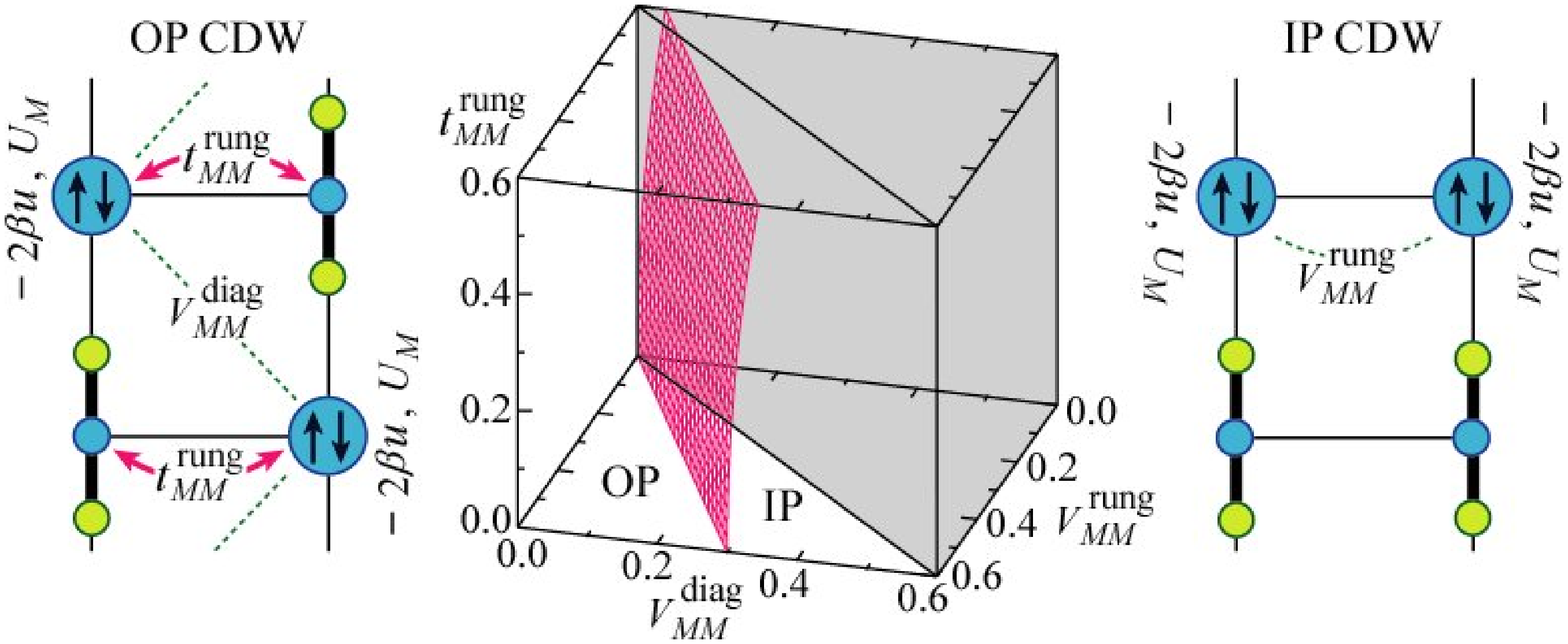}
\vspace*{-2mm}
\caption{(Color online)
         Hartree-Fock calculation of a ground-state phase diagram on the
         $t_{M\!M}^{\rm rung}$-$V_{M\!M}^{\rm rung}$-$V_{M\!M}^{\rm diag}$
         cube within a single-band model, where $t_{M\!M}^{\rm leg}$ is
         taken as unity and the region of no physical interest is shaded.
         A simple consideration of competing IP-CDW and OP-CDW states is
         also presented, where Coulomb energy losses and transfer energy
         gains within $d$ electrons are illustrated.}
\label{F:PhD}
%\end{figure}
%\begin{figure}
\vspace*{3mm}
\centering
\includegraphics[width=85mm]{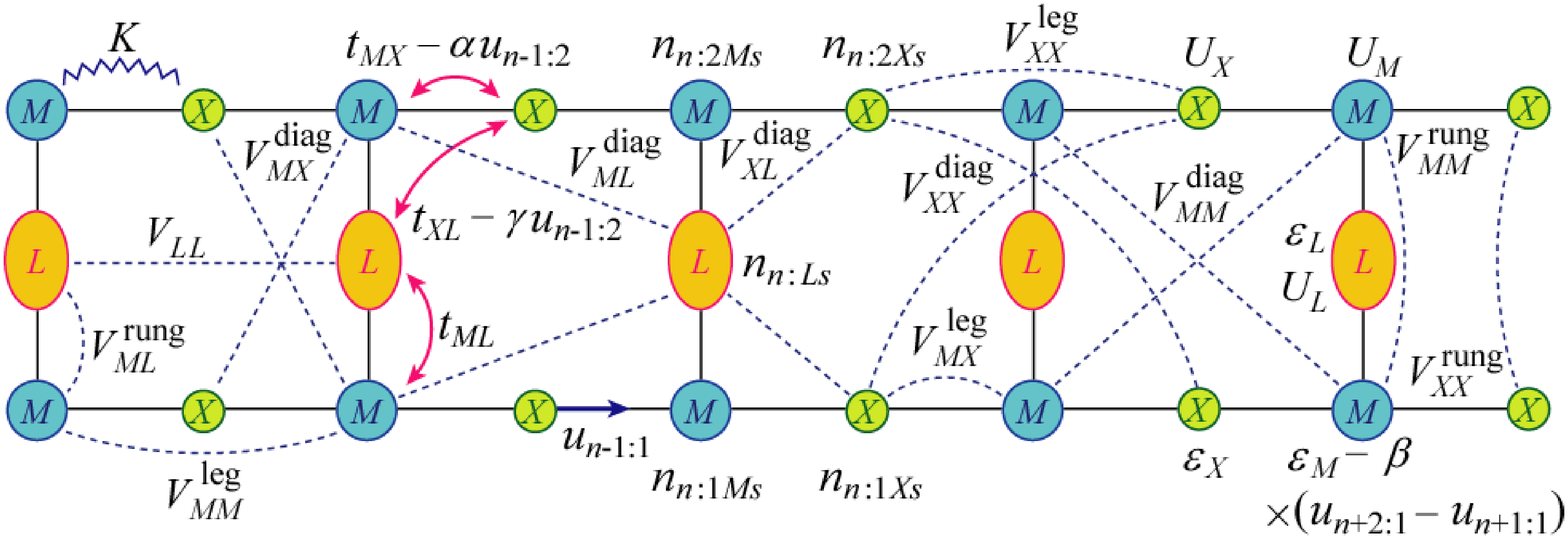}
\vspace*{-2mm}
\caption{(Color online)
         Modeling of $M\!X$ ladders, where $M=\mbox{Pt}$;
         $X=\mbox{Br},\mbox{Cl}$; $L=\mu\mbox{-bpym},\mbox{bpy}$.
         An electron with spin $s=\uparrow,\downarrow\equiv\pm$ is created
         on the $M\,d_{z^2}$ and $X\,p_z$ orbitals on the $l$th leg in the
         $n$th unit by $a_{n:lMs}^\dagger$ and $a_{n:lXs}^\dagger$,
         respectively, and on the $n$th-rung $L\,\pi$ orbital by
         $a_{n:Ls}^\dagger$.
         The resultant electron density is given by
         $a_{n:lMs}^\dagger a_{n:lMs}\equiv n_{n:lMs}$,
         $a_{n:lXs}^\dagger a_{n:lXs}\equiv n_{n:lXs}$, and
         $a_{n:Ls}^\dagger a_{n:Ls}\equiv n_{n:Ls}$.
         The on-site energies (electron affinities) of isolated atoms and
         molecules are given by $\varepsilon_A$ ($A=M,X,L$) and the
         electron hoppings between these levels are modeled by $t_{A\!A'}$
         [$A=M,X$; $A'(\neq A)=X,L$].
         $U_A$ and $V_{A\!A'}$ ($A,A'=M,X,L$) describe the on-site and
         different-site Coulomb interactions, respectively.
         The halogen-ion displacements $u_{n:l}$ affect electrons through
         intersite ($\alpha,\gamma$) and intrasite ($\beta$) coupling
         constants, accompanied by elastic energy $\propto K$.}
\label{F:H}
\end{figure}

\section{Ground-State Phase Competition}

   Resonant Raman spectra of (bpym)[Pt(en)$X$]$_2$ and
(bpy)[Pt(dien)Br]$_2$ both suggest a Pt$^{2+}$/Pt$^{4+}$
[Pt$^{3\mp\delta}$ ($0<\delta\leq 1$) in practice]
valence-alternating ground state. \cite{K12066,K7372}
Then how is the interchain valence arrangement,
in phase (IP) or out of phase (OP)?
CDW states of the IP and OP types are indeed in close competition with
varying interchain electronic communication.

   Let us consider a half-filled single-band Hamiltonian,
\begin{eqnarray}
   &&
   {\cal H}=
   \sum_{n=1}^N\sum_{l=1}^2
   \biggl\{
    Ku_{n:l}^2
   -\sum_{s=\pm}
    \Bigl[
     \beta\big(u_{n:l}-u_{n-1:l}\big)n_{n:lMs}
   \nonumber \\
   &&\qquad
    +\Bigl(t_{M\!M}^{\rm leg}a_{n+1:lMs}^\dagger a_{n:lMs}
     +\frac{t_{M\!M}^{\rm rung}}{2}a_{n:lMs}^\dagger a_{n:3-lMs}
   \nonumber \\
   &&\qquad
     +{\rm H.c.}
     \Bigr)
    \Bigr]
   +\sum_{s,s'=\pm}
    \Bigl(
     \frac{\delta_{-s,s'}}{2}U_M n_{n:lMs}n_{n:lMs'}
   \nonumber \\
   &&\qquad
    +V_{M\!M}^{\rm leg} n_{n:lMs}n_{n+1:lMs'}
    +\frac{V_{M\!M}^{\rm rung}}{2} n_{n:lMs}n_{n:3-lMs'}
   \nonumber \\
   &&\qquad
    +V_{M\!M}^{\rm diag} n_{n:lMs}n_{n+1:3-lMs'}
    \Bigr)
   \biggr\},
   \label{E:HSB}
\end{eqnarray}
assuming the halogen $p_z$ and ligand $\pi$ orbitals to be fully filled
and thus inactive.
Here, except for the intrachain and interchain metal-to-metal
supertransfers, $t_{M\!M}^{\rm leg}$ and $t_{M\!M}^{\rm rung}$, we use
the same notation that is later defined in Eq. (\ref{E:H}) and
Fig. \ref{F:H}.
The second-order perturbation scheme under the conditions of
$t_{M\!M}^{\rm rung}\ll t_{M\!M}^{\rm leg}$ gives the energies
of IP- and OP-CDW states as
\begin{eqnarray}
   &&
   \frac{E_{\rm IP}}{N}
  =2Ku^2-4\beta u+U_M+2V_{M\!M}^{\rm rung},
   \\
   &&
   \frac{E_{\rm OP}}{N}
  =2Ku^2-4\beta u+U_M+4V_{M\!M}^{\rm diag}
   \nonumber \\
   &&\qquad
  -\frac{2(t_{M\!M}^{\rm rung})^2}
        {4\beta u-U_M+V_{M\!M}^{\rm rung}
        +4(V_{M\!M}^{\rm leg}-V_{M\!M}^{\rm diag})},\qquad
\end{eqnarray}
and they are balanced at
\begin{eqnarray}
   &&
   (t_{M\!M}^{\rm rung})^2
  =(2V_{M\!M}^{\rm diag}-V_{M\!M}^{\rm rung})
   \nonumber \\
   &&\qquad\times
   [4\beta u-U_M+V_{M\!M}^{\rm rung}
   +4(V_{M\!M}^{\rm leg}-V_{M\!M}^{\rm diag})],\qquad
   \label{E:PhD}
\end{eqnarray}
where $u\equiv|u_{n:l}|$ is the halogen-ion displacement in isolated
$M\!X$ chains.
We show in Fig. \ref{F:PhD} a numerical phase diagram based on the
Hamiltonian (\ref{E:HSB}), which agrees well to the estimate
(\ref{E:PhD}).
IP CDW and OP CDW are stabilized with increasing $V_{M\!M}^{\rm diag}$ and
$V_{M\!M}^{\rm rung}$, respectively.
OP CDW is further stabilized with increasing $t_{M\!M}^{\rm rung}$, while
IP CDW has no chance of interchain electron transfer without $\pi$
orbitals mediation (in the strongly valence-trapped limit, strictly).
Nonvanishing optical absorption in the rung direction with the IP-CDW
background should be significant of contributive ligand $\pi$ orbitals.

   While the ground-state phase diagram remains almost unchanged with $p$
and/or $\pi$ electrons taken into calculation, the single-band model
totally fails to interpret the optical properties.
The optical conductivity spectra measured on (bpym)[Pt(en)$X$]$_2$ and
(bpy)[Pt(dien)Br]$_2$ are considerably different from each other, but it
cannot distinguish between them at all.
We proceed to much more elaborate calculations.
\begin{table*}
\caption{Model parameters for
         (bpym)[Pt(en)Cl]$_2$ (A),
         (bpym)[Pt(en)Br]$_2$ (B), and
         (bpy)[Pt(dien)Br]$_2$ (C),
         where $\varepsilon_M$ is set equal to zero.}
\begin{ruledtabular}
\begin{tabular}{lrrrlrrrlrrrlrrr}
 $$ & A\ $\,$ & B\ $\,$ & C\ $\,$ &
 $$ & A\ $\,$ & B\ $\,$ & C\ $\,$ &
 $$ & A\ $\,$ & B\ $\,$ & C\ $\,$ &
 $$ & A\ $\,$ & B\ $\,$ & C\ $\,$ \\
\hline
 $t_{M\!X}\,$(eV)$\!\!$ & $1.54$ & $1.35$ & $1.32$ &
 $V_{M\!X}^{\rm leg}\,$(eV)$\!\!$  & $0.80$ & $0.92$ & $0.86$ &
 $V_{M\!X}^{\rm diag}\,$(eV)$\!\!$ & $0.22$ & $0.19$ & $0.14$ &
 $\varepsilon_X\,$(eV)$\!\!\!\!$ & $-3.39$ & $-2.43$ & $-2.38$ \\
 $t_{M\!L}\,$(eV)$\!\!$ & $1.08$ & $0.94$ & $0.53$ &
 $V_{M\!M}^{\rm leg}\,$(eV)$\!\!$  & $0.23$ & $0.20$ & $0.20$ &
 $V_{M\!M}^{\rm diag}\,$(eV)$\!\!$ & $0.20$ & $0.16$ & $0.07$ &
 $\varepsilon_L\,$(eV)$\!\!\!\!$ & $-0.62$ & $-0.54$ & $-0.66$ \\
 $t_{X\!L}\,$(eV)$\!\!$ & $0.15$ & $0.13$ & $0.05$ &
 $V_{X\!X}^{\rm leg}\,$(eV)$\!\!$  & $0.18$ & $0.16$ & $0.14$ &
 $V_{X\!X}^{\rm diag}\,$(eV)$\!\!$ & $0.14$ & $0.12$ & $0.08$ &
 $\beta\,$(eV/\AA)$\!\!\!\!$ & $2.37$ & $2.40$ & $2.27$ \\
 $U_M\,$(eV)$\!\!$ & $1.23$ & $1.08$ & $0.92$ &
 $V_{M\!L}^{\rm rung}\,$(eV)$\!\!$ & $0.38$ & $0.35$ & $0.30$ &
 $V_{M\!L}^{\rm diag}\,$(eV)$\!\!$ & $0.18$ & $0.13$ & $0.08$ &
 $\alpha\,$(eV/\AA)$\!\!\!\!$ & $0.85$ & $0.77$ & $0.65$ \\
 $U_X\,$(eV)$\!\!$ & $1.69$ & $0.94$ & $1.06$ &
 $V_{M\!M}^{\rm rung}\,$(eV)$\!\!$ & $0.23$ & $0.20$ & $0.16$ &
 $V_{X\!L}^{\rm diag}\,$(eV)$\!\!$ & $0.25$ & $0.23$ & $0.20$ &
 $\gamma\,$(eV/\AA)$\!\!\!\!$& $0.30$ & $0.26$ & $0.13$ \\
 $U_L\,$(eV)$\!\!$ & $1.08$ & $0.94$ & $1.06$ &
 $V_{X\!X}^{\rm rung}\,$(eV)$\!\!$ & $0.18$ & $0.16$ & $0.13$ &
 $V_{L\!L}\,$(eV)$\!\!$            & $0.23$ & $0.19$ & $0.17$ &
 $K\,$(eV/\AA$^2$)$\!\!\!\!$& $6.00$ & $8.00$ & $8.00$ \\
\end{tabular}
\end{ruledtabular}
\label{T:MP}
\end{table*}

\section{Model Hamiltonian}

   We consider a multiband extended Peierls-Hubbard Hamiltonian of
$4/5$ electron filling on the two-leg ladder lattice,
\begin{widetext}
\begin{eqnarray}
   &&
   {\cal H}=
   \sum_{n,l,s}
   \Bigl\{
    \big[\varepsilon_{M}-\beta\big(u_{n:l}-u_{n-1:l}\big)\big]n_{n:lMs}
   +\varepsilon_{X}n_{n:lXs}
   +\frac{\varepsilon_{L}}{2}n_{n:Ls}
   \Bigr\}
  -\sum_{n,l,s}
   \Bigl[
    \big(t_{M\!X}+\alpha u_{n:l}\big)
    a_{n+1:lMs}^\dagger a_{n:lXs}
   \nonumber \\
   &&\qquad
   +\big(t_{M\!X}-\alpha u_{n:l}\big)
    a_{n:lXs}^\dagger a_{n:lMs}
   +t_{M\!L}a_{n:lMs}^\dagger a_{n:Ls}
   +\big(t_{X\!L}+\gamma u_{n:l}\big)
    a_{n+1:Ls}^\dagger a_{n:Xs}
   +\big(t_{X\!L}-\gamma u_{n:l}\big)
    a_{n:lXs}^\dagger a_{n:Ls}
   \nonumber \\
   &&\qquad
   +{\rm H.c.}
   \Bigr]
  +\sum_{n,l,s,s'}
   \biggl\{
    \frac{\delta_{-s,s'}}{2}
    \biggl(
     \sum_{A=M,X}
     U_{A}n_{n:lAs}n_{n:lAs'}
    +\frac{U_{L}}{2}n_{n:Ls}n_{n:Ls'}
    \biggr)
   +V_{M\!X}^{\rm leg}\big(n_{n+1:lMs}+n_{n:lMs}\big)n_{n:lXs'}
   \nonumber \\
   &&\qquad
   +V_{M\!L}^{\rm rung}n_{n:lMs}n_{n:Ls'}
   +V_{X\!L}^{\rm diag}n_{n:lXs}\big(n_{n:Ls'}+n_{n+1:Ls'}\big)
   +V_{M\!L}^{\rm diag}\big(n_{n+1:lMs}n_{n:Ls'}+n_{n:lMs}n_{n+1:Ls'}\big)
   \nonumber \\
   &&\qquad
   +V_{M\!X}^{\rm diag}\big(n_{n+1:lMs}+n_{n:lMs}\big)n_{n:3-lXs'}
  +\sum_{A=M,X}
   \Bigl(
    V_{A\!A}^{\rm leg}n_{n+1:lAs}n_{n:lAs'}
   +\frac{V_{A\!A}^{\rm rung}}{2}n_{n:lAs}n_{n:3-lAs'}
   \nonumber \\
   &&\qquad
   +V_{A\!A}^{\rm diag}n_{n+1:lAs}n_{n:3-lAs'}
   \Bigr)
   +\frac{V_{L\!L}}{2} n_{n+1:Ls}n_{n:Ls'}
   \biggr\}
  +K\sum_{n,l} u_{n:l}^{2},
   \label{E:H}
\end{eqnarray}
\end{widetext}
as is illustrated with Fig. \ref{F:H}, where platinum $d_{z^2}$,
halogen $p_z$, and rung-ligand $\pi$ orbitals are explicitly taken into
calculation.
For platinum-halide single-chain compounds, typical Pt-$X$ stretching modes
are observed with frequencies of $10$ to $40\,\mbox{meV}$ and their
coupling strength is estimated to be $2$ to $3\,\mbox{eV/\AA}$.
\cite{D3285}
Assuming the in-chain vibrational modes to remain valid in our ladder
materials, the phonon energy is less than one percent of the optical gap
$E_{\rm CT}$ and at most four percent of the electron-phonon interaction
$\beta u$ (see Tables \ref{T:MP} and \ref{T:LC} later on).
That is why we stand on the adiabatic Hamiltonian (\ref{E:H}).
Such a classical treatment of phonons is widely adopted and generally
successful for mixed-valent platinum-halide compounds.
\cite{G6408,Y125124,K2163,F044717,N3865,B339,T1800,Y165113}
Quantum phonons may be relatively effective in strongly-correlated
valence-delocalized nickel-halide chains. \cite{W6435}

   Characterization of the brandnew $M\!X$ ladders is still in the early
stage from both experimental \cite{K12066,K7372} and theoretical
\cite{F044717,I063708} points of view, and therefore, little is known
about the model parameters.
In such circumstances, extensive two-band model studies on $M\!X$ single
chains serve as guides to our exploration.

   Since the hopping integral $t_{M\!X}$ is particularly important as an
energy scale, we first set it closely consistent with the authorized
estimates obtained by the Los Alamos National Laboratory working team.
Comparing two-band model descriptions with first-principle
local-density-approximation calculations, they report that
$t_{\rm PtCl}=1.54\,\mbox{eV}$ and $t_{\rm PtBr}=1.30\,\mbox{eV}$ for
[Pt(en)$_2X$], \cite{W6435,A169,A2739} while
$t_{\rm PtCl}=1.60\,\mbox{eV}$ and $t_{\rm PtBr}=1.50\,\mbox{eV}$ for
[Pt(NH$_3$)$_2X_3$]. \cite{A3104,A1415}
Here we take $t_{M\!X}$ to be $1.54\,\mbox{eV}$, $1.35\,\mbox{eV}$, and
$1.32\,\mbox{eV}$ for (bpym)[Pt(en)Cl]$_2$, (bpym)[Pt(en)Br]$_2$, and
(bpy)[Pt(dien)Br]$_2$, respectively, considering the consistency of the
resultant theoretical findings with experimental observations.

   Another essential one-body parameter, the relative on-site energy
$\varepsilon_M-\varepsilon_X$, may also be less dependent on the rung
ligands, but it is not so established as $t_{M\!X}$ even in $M\!X$ single
chains.
The Los Alamos group on one hand reports that
$\varepsilon_{\rm Cl}=-1.32\,\mbox{eV}$ and
$\varepsilon_{\rm Br}=-0.58\,\mbox{eV}$ for
[Pt(en)$_2X$], \cite{W6435,A169,A2739} while
$\varepsilon_{\rm Cl}=-2.90\,\mbox{eV}$ and
$\varepsilon_{\rm Br}=-2.30\,\mbox{eV}$ for
[Pt(NH$_3$)$_2X_3$], \cite{A3104,A1415} but on the other hand suggests
another possibility that
$\varepsilon_{\rm Cl}=-4.24\,\mbox{eV}$ and
$\varepsilon_{\rm Br}=-1.20\,\mbox{eV}$ for
[Pt(en)$_2X$], \cite{S1659} where $\varepsilon_M$ is set equal to zero.
Therefore, we tune the on-site energies within these estimates so as to
reproduce experimental observations.

   Coulomb interactions much more vary with the surrounding ligands and
seriously depend on the modeling.
For instance, the on-site repulsion $U_{\rm Pt}$ effectively amounts to
a few eV in a pure Hubbard model, \cite{S1659,C723} whereas it is strongly
suppressed to a half eV or less in a fully extended model with power-law
decaying Coulomb terms. \cite{B6065}
Taking it into consideration that any empirical estimate of $U_{\rm Pt}$
does not exceed $2\,\mbox{eV}$, \cite{H5706,K18682,K1789,W3013,S3066}
relying upon a well established criterion
$U_{\rm Pt}\simeq U_{\rm Br}\alt U_{\rm Cl}$, \cite{W6435,A2739}
and strictly keeping the restriction that the farther, the smaller,
we compare our calculations with experimental findings on an absolute
scale.
The thus-obtained $d$-$p$-$\pi$ model parameters for $M\!X$ ladders are
listed in Table \ref{T:MP}.
Among the Coulomb correlation parameters employed,
$U_A$ ($A=M,X,L$) and $V_{M\!X}^{\rm leg}$ play predominant roles in
reproducing main features of the optical conductivity spectra.
The rest are much less effective for the optical properties.
Indeed we have many parameters, but their output is not so adjustable as
might be expected.
The effect and tuning of each parameter is further discussed and
visualized in Appendix \ref{A:tuning} in order to demonstrate the
reliability of our parametrization.
\begin{table}
\vspace*{-2mm}
\caption{Theoretical (bare) and experimental (parenthesized) estimates
         of structural and optical parameters for
         (bpym)[Pt(en)Cl]$_2$ (A),
         (bpym)[Pt(en)Br]$_2$ (B), and
         (bpy)[Pt(dien)Br]$_2$ (C).
         $2c_{M\!X}$ and $2c_{M\!L}$ correspond to the Pt-$X$-Pt and
         Pt-$L$-Pt distances, respectively, while $u$ to the $X$
         displacement from the mid point.
         $E_{\rm CT}$ is the intrachain charge-transfer excitation
         energy.}
\begin{ruledtabular}
\begin{tabular}{ccccc}
\noalign{\vspace*{-0.5mm}}
                  &
$c_{M\!X}\,$(\AA) &
$c_{M\!L}\,$(\AA) &
$u\,$(\AA)        &
$E_{\rm CT}\,$(eV)\\
\noalign{\vspace*{0.5mm}}
\hline
A &$1.98$ ($2.72$)&$1.98$ ($2.73$)&$0.38$ ($0.40$)&$3.74$ ($3.66$)\\
B &$2.00$ ($2.77$)&$2.00$ ($2.74$)&$0.28$ ($0.29$)&$2.47$ ($2.36$)\\
C &$3.50$ ($2.73$)&$7.00$ ($5.59$)&$0.25$ ($0.23$)&$2.13$ ($2.18$)\\
\end{tabular}
\end{ruledtabular}
\label{T:LC}
\end{table}

   Table \ref{T:LC} claims that our theory well interprets X-ray
diffraction measurements as well as optical observations.
\cite{K12066,K7372}
Since optical conductivity of definite polarization is proportional to the
relevant interatomic separation squared [see Eqs. (\ref{E:Jparallel}) and
(\ref{E:Jperp})], a direct comparison of the calculations to the bare
observations gives lattice constants.
The Peierls gap and the lattice distortion, which are in proportion to
each other, are determined within our calculation independent of any
optical measurement.
The optical excitation energy $E_{\rm CT}$ is closely related but does not
coincide with the Peierls gap in the present case.
The consequent lattice parameters correspond to the observations within
a factor $1.4$, which guarantees our interpretation of the optical
conductivity spectra.
The calculated optical gaps are also in good agreement with the
observations, which justifies our parametrization.
The on-site repulsion $U_M$ and the site-diagonal coupling constant
$\beta$ competitively dominate $E_{\rm CT}$, whereas the elastic constant
$K$ is decisive of $u$.
A general tendency for halogen-ion displacements,
$u({\rm I})<u({\rm Br})<u({\rm Cl})$, holds in $M\!X$ ladders \cite{K7372}
as well as in $M\!X$ single chains. \cite{O2023}

\section{Optical Conductivity Spectra}
\subsection{Calculational Procedure}

\begin{figure}
\centering
\includegraphics[width=84mm]{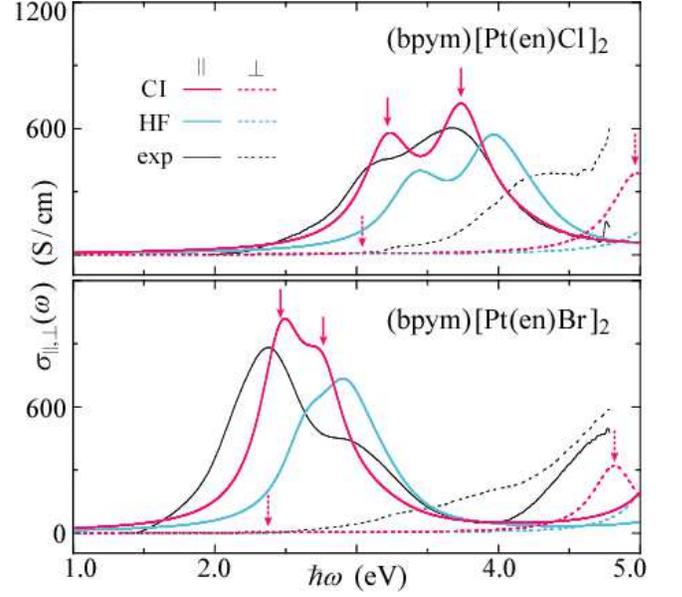}
\vspace*{-2mm}
\caption{(Color online)
         Hartree-Fock (HF) and single-excitation configuration-interaction
         (CI) calculations of the polarized optical conductivity spectra
         parallel ($\parallel$) and perpendicular ($\perp$) to ladder legs
         for IP-CDW states in comparison with experimental observations
         (exp) of (bpym)[Pt(en)$X$]$_2$.}
\label{F:OCIP}
\end{figure}

\begin{figure}
\centering
\includegraphics[width=84mm]{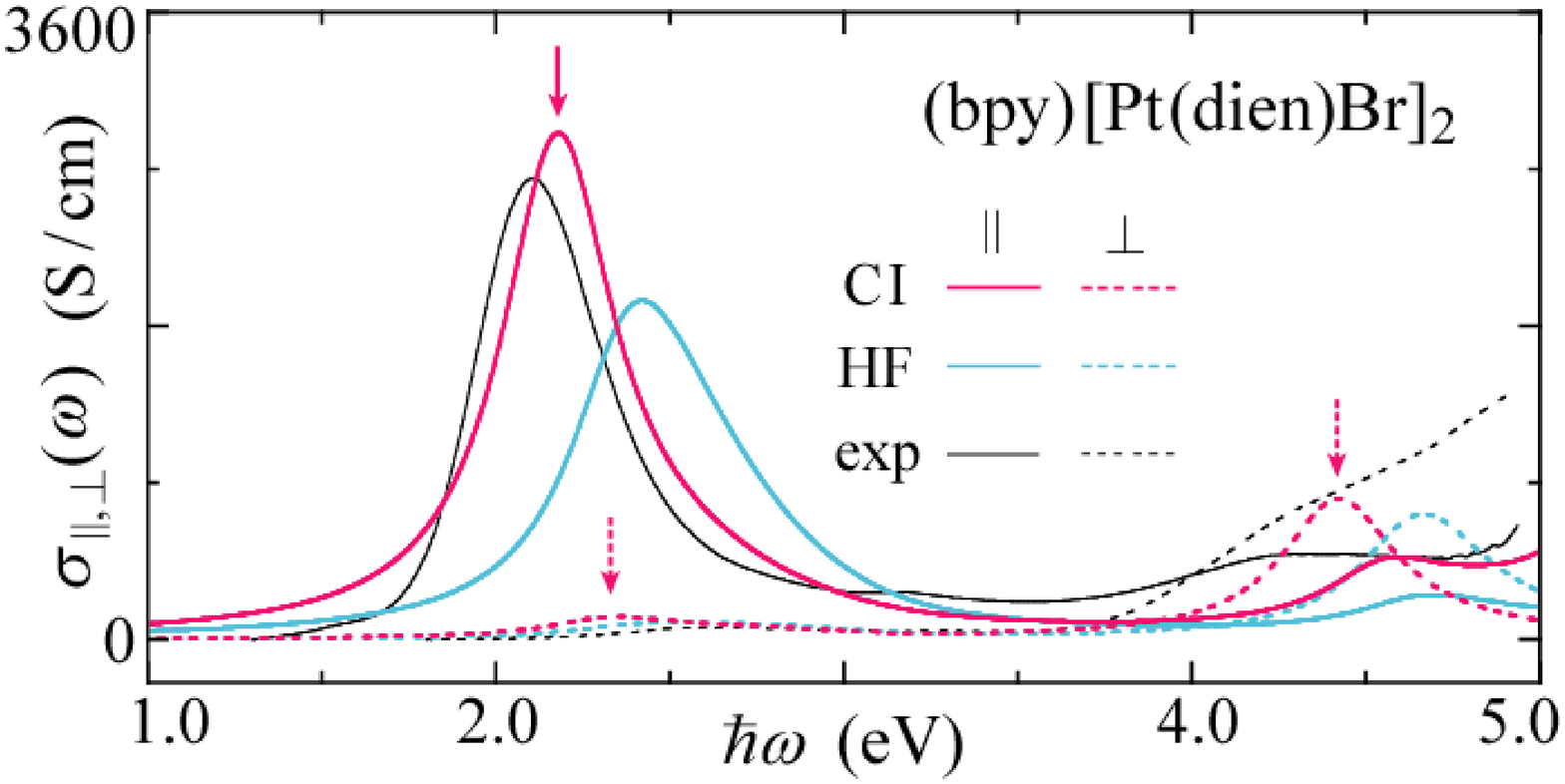}
\vspace*{-2mm}
\caption{(Color online)
         Hartree-Fock (HF) and single-excitation configuration-interaction
         (CI) calculations of the polarized optical conductivity spectra
         parallel ($\parallel$) and perpendicular ($\perp$) to ladder legs
         for an OP-CDW state in comparison with experimental observations
         (exp) of (bpy)[Pt(dien)Br]$_2$.}
\label{F:OCOP}
\end{figure}

   In order to discuss optical absorption as a function of the
polarization of incident light ($\mbox{\boldmath$E$}_{\rm in}$), we define
current operators along ladder legs ($\parallel\mbox{\boldmath$c$}$) and
rungs ($\perp\mbox{\boldmath$c$}$) as
\begin{widetext}
\begin{eqnarray}
   &&
   {\cal J}_\parallel
   =\frac{{\rm i}e}{\hbar}c_{M\!X}\sum_{l,n,s}
    \bigl[
     (t_{M\!X}+\alpha u_{n:l})
     a_{n+1:lMs}^\dagger a_{n:lXs}
    +(t_{M\!X}-\alpha u_{n:l})
     a_{n:lXs}^\dagger a_{n:lMs}
   \nonumber \\
   &&\qquad
    +(t_{X\!L}+\gamma u_{n:l})
     a_{n+1:Ls}^\dagger a_{n:lXs}
    +(t_{X\!L}-\gamma u_{n:l})
     a_{n:lXs}^\dagger a_{n:Ls}
    -{\rm H.c.}
    \bigr],
   \label{E:Jparallel}
   \\
   &&
   {\cal J}_\perp
   =\frac{{\rm i}e}{\hbar}c_{M\!L}\sum_{l,n,s}(-1)^l
    \bigl[
     t_{M\!L} a_{n:lMs}^\dagger a_{n:Ls}
    +(t_{X\!L}-\gamma u_{n:l})
     a_{n:lXs}^\dagger a_{n:Ls}
    +(t_{X\!L}+\gamma u_{n:l})
     a_{n:lXs}^\dagger a_{n+1:Ls}
    -{\rm H.c.}
    \bigr],\qquad
   \label{E:Jperp}
\end{eqnarray}
\end{widetext}
where $2c_{M\!X}$ and $2c_{M\!L}$ are the intermetallic separations in the
leg and rung directions, respectively, and are fixed at
$c_{M\!X}=c_{M\!L}$ and $2c_{M\!X}=c_{M\!L}$ for
(bpym)[Pt(en)$X$]$_2$ and (bpy)[Pt(dien)Br]$_2$, respectively,
in our calculation.
Since the charge-transfer excitation energy is of eV order,
\cite{K12066,K7372} the system effectively lies in the ground state at
room temperature.
Then the real part of the optical conductivity reads
\begin{equation}
   \sigma_{\parallel,\perp}(\omega)
    =\frac{\pi}{\omega}\sum_i
   |\langle E_i|{\cal J}_{\parallel,\perp}|E_0\rangle|^2
   \delta(E_i-E_0-\hbar\omega),
   \label{E:OC}
\end{equation}
where $|E_i\rangle$ is the $i$th-lying state of energy $E_i$.
$|E_0\rangle$ is defined as
\begin{equation}
   |E_0\rangle
  =\prod_{\epsilon_{\mu s}\leq\epsilon_{\rm F}}
   c_{\mu+}^\dagger c_{\mu-}^\dagger|0\rangle,
\end{equation}
where $|0\rangle$ is the true electron vacuum, $\epsilon_{\rm F}$ is the
Fermi energy, and $c_{\mu s}^\dagger$ creates an electron of spin $s$
in the Hartree-Fock (HF) eigenstate with an eigenvalue $\epsilon_{\mu s}$.
Excited states are calculated within and beyond the HF scheme, being
generally defined as
\begin{equation}
   |E_i\rangle
   =\sum_{\epsilon_{\mu s}\leq\epsilon_{\rm F}<\epsilon_{\nu s}}
    f(\mu,\nu,s;i)c_{\nu s}^\dagger c_{\mu s}|E_0\rangle.
\end{equation}
Every excited state of the HF type is a single Slater determinant, where
$f(\mu,\nu,s;i)=\delta_{\mu\nu s,i}$.
Those of the configuration-interaction (CI) type consist of
resonating Slater determinants, where $f(\mu,\nu,s;i)$ satisfies
\begin{eqnarray}
   &&
   \sum_{\epsilon_{\mu s}\leq\epsilon_{\rm F}<\epsilon_{\nu s}}
   \langle E_0|
    c_{\mu's'}^\dagger c_{\nu's'}{\cal H}c_{\nu s}^\dagger c_{\mu s}
   |E_0\rangle
   f(\mu,\nu,s;i)
   \nonumber \\
   &&\qquad\qquad
  =E_i f(\mu',\nu',s';i),
\end{eqnarray}
that is, the unitary matrix $f(\mu,\nu,s;i)$ diagonalizes
the original Hamiltonian ${\cal H}$.
Since the HF Hamiltonian ${\cal H}_{\rm HF}$ is diagonal with respect to
pure particle-hole states as
$\langle E_0|
  c_{\mu's'}^\dagger c_{\nu's'}
  {\cal H}_{\rm HF}
  c_{\nu s}^\dagger c_{\mu s}
 |E_0\rangle
=\delta_{\mu'\nu's',\mu\nu s}(E_0-\epsilon_{\mu s}+\epsilon_{\nu s})$,
the residual component ${\cal H}-{\cal H}_{\rm HF}$ mixes the Slater
determinants and reduces the interband transition energy
(see Figs. \ref{F:OCIP} and \ref{F:OCOP} later).
Equation (\ref{E:OC}) calculated is Lorentzian broadened.
\begin{figure}
\centering
\includegraphics[width=84mm]{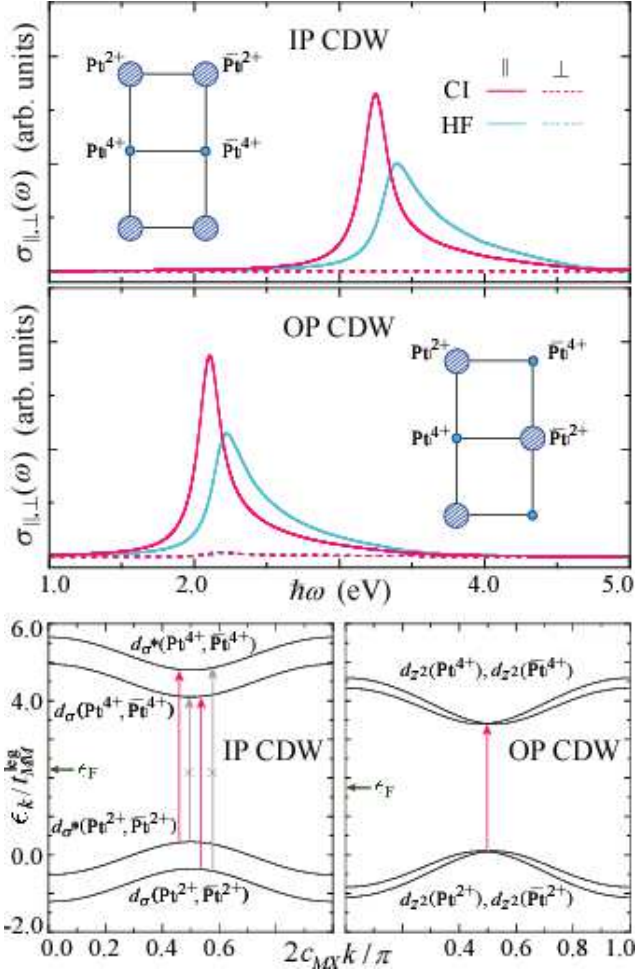}
\vspace*{-2mm}
\caption{(Color online)
         Hartree-Fock (HF) and single-excitation configuration-interaction
         (CI) calculations of the polarized optical conductivity spectra
         parallel ($\parallel$) and perpendicular ($\perp$) to ladder
         legs within the single-band model, where
         $t_{M\!M}^{\rm leg}=0.75\,\mbox{eV}\ (0.66\,\mbox{eV})$,
         $t_{M\!M}^{\rm rung}/t_{M\!M}^{\rm leg}=0.35\ (0.16)$,
         $U_M/t_{M\!M}^{\rm leg}=1.40\ (1.20)$,
         $V_{M\!M}^{\rm leg}/t_{M\!M}^{\rm leg}=0.35\ (0.35)$,
         $V_{M\!M}^{\rm rung}/t_{M\!M}^{\rm leg}=0.30\ (0.25)$,
         $V_{M\!M}^{\rm diag}/t_{M\!M}^{\rm leg}=0.26\ (0.10)$, and
         $\beta/\sqrt{t_{M\!M}^{\rm leg}K}=1.10\ (1.00)$
         for IP CDW (OP CDW).
         Hartree-Fock calculations of the relevant dispersion relations
         are also shown in an attempt to understand the spectral features
         of $\sigma_\parallel(\omega)$, where bare arrows and those with a
         cross attached denote major optical absorptions and optically
         forbidden transitions, respectively.}
\label{F:OCSB}
\end{figure}

\subsection{Single-Band Calculation}

   Before analyzing experimental findings in detail, we calculate the
optical conductivity in terms of the single-band Hamiltonian (\ref{E:HSB})
in an attempt to demonstrate the indispensable halogen $p_z$ and ligand
$\pi$ orbitals.
Figure \ref{F:OCSB} shows that the optical observations of IP- and OP-CDW
states are quite alike without contributive $p$ and $\pi$ electrons.
The spectra in the leg direction are single-peaked, whereas no significant
spectral weight lies in the rung direction.

   Such observations are well understandable when we consider the
underlying energy structures.
Since the $d_{z^2}$ orbitals of equally valent platinum ions have the same
energy and are well hybridized with each other, we find well split
filled/conduction bands in IP CDW, while the two intrachain
$d_{z^2}(\mbox{Pt}^{2+})$/$d_{z^2}(\mbox{Pt}^{4+})$ bands remain almost
degenerate with each other in OP CDW.
The pronounced peak of $\sigma_\parallel(\omega)$ is attributed to the
interband excitations at the zone center.
When an electron is pumped up from the filled to conduction bands, there
are four types of transitions possible in general.
However, the lowest- and highest-energy ones are optically forbidden and
the rest, optically allowed, cost the same energy.
The conduction and filled bands are exactly symmetric with respect to the
Fermi level due to the electron-hole symmetry preserved.
That is why not only the OP-CDW spectrum but also the IP-CDW spectrum is
single-peaked.
The single-peak structure remains unchanged with excitonic effect on.
A consideration of $X$ $p_z$ and/or $L$ $\pi$ orbitals leads to the
breakdown of the electron-hole symmetry and lifts the degeneracy between
the optical observations of IP CDW and OP CDW, which is essential to the
understanding of experimental findings.

   The vanishing weight of $\sigma_\perp(\omega)$ is also due to the
sleeping $p$ and $\pi$ electrons and is never in agreement with any
experiment.
We are thus lead to the $d$-$p$-$\pi$ modeling.

\begin{figure*}
\centering
\includegraphics[width=168mm]{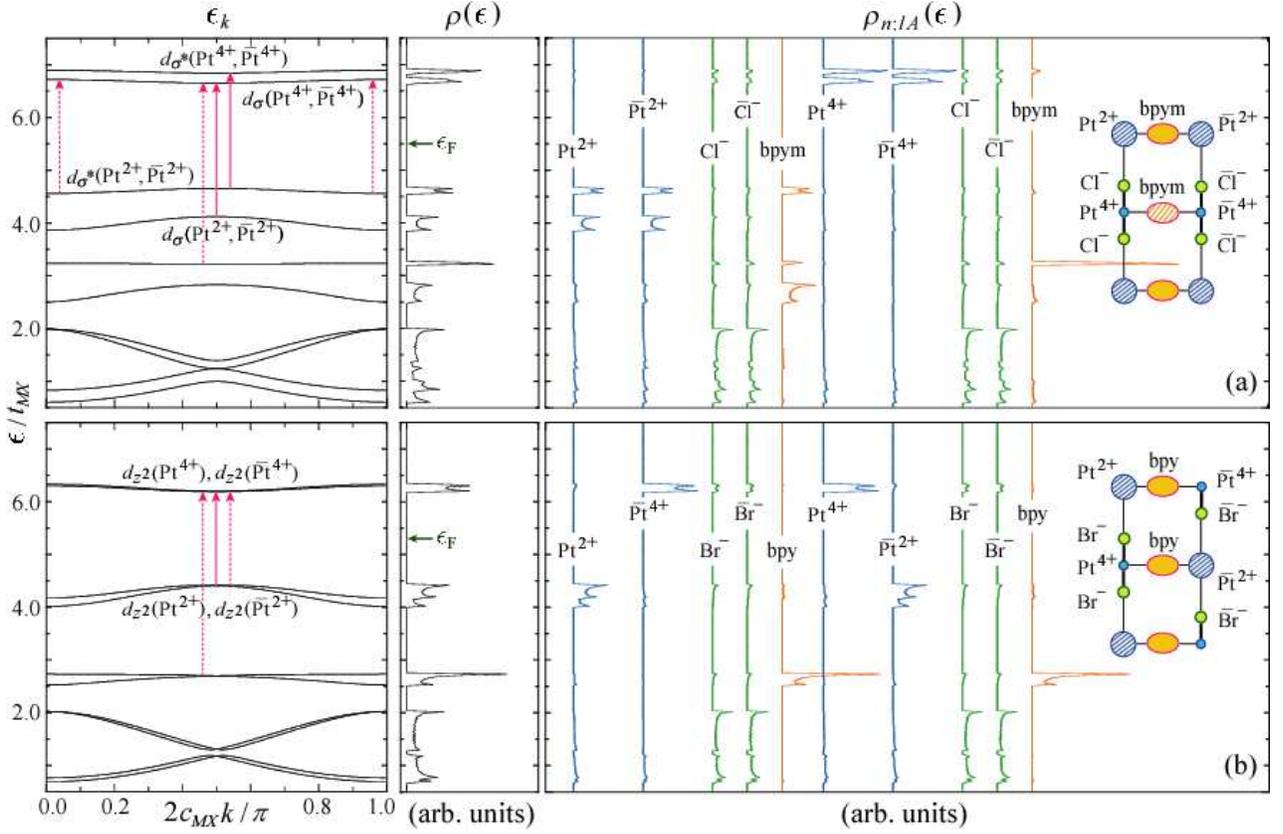}
\vspace*{-2mm}
\caption{(Color online)
         Hartree-Fock calculations of the dispersion relation $\epsilon_k$
         and the local (total) density of states $\rho_{n:lA}(\epsilon)$
         [$\rho(\epsilon)\equiv\sum_{n,l,A}\rho_{n:lA}(\epsilon)$] for the
         IP-CDW (a) and OP-CDW (b) states describing (bpym)[Pt(en)Cl]$_2$
         and (bpy)[Pt(dien)Br]$_2$, respectively.
         Major optical absorptions with
         $\mbox{\boldmath$E$}_{\rm in}\parallel\mbox{\boldmath$c$}$ and
         $\mbox{\boldmath$E$}_{\rm in}\perp\mbox{\boldmath$c$}$ are
         indicated by solid and dotted arrows, respectively,
         in Figs. \ref{F:OCIP} and \ref{F:OCOP} as well as here.}
\label{F:rho}
\vspace*{-2mm}
\end{figure*}

\subsection
{$\mbox{\boldmath$d$}$-$\mbox{\boldmath$p$}$-$\mbox{\boldmath$\pi$}$
 Description}

   In Figs. \ref{F:OCIP} and \ref{F:OCOP} we compare the $d$-$p$-$\pi$
calculations of the optical conductivity with experimental observations,
\cite{K7372} that is, the Kramers-Kronig transforms of polarized
reflectivity spectra for the single crystals at room temperature.
The calculations qualitatively interpret most of the spectral features
within the HF scheme and quantitatively improves with excitonic effects.
We have two arguments in particular:
i) For $\mbox{\boldmath$E$}_{\rm in}\parallel\mbox{\boldmath$c$}$,
the main absorption band is double-peaked in (bpym)[Pt(en)$X$]$_2$ but
single-peaked in (bpy)[Pt(dien)Br]$_2$;
ii) For $\mbox{\boldmath$E$}_{\rm in}\perp\mbox{\boldmath$c$}$,
significant absorption is observed not only in (bpy)[Pt(dien)Br]$_2$ but
also in (bpym)[Pt(en)$X$]$_2$.

   $\sigma_\parallel(\omega)$---Corresponding spectra measured on the
$M\!X$ single-chain compounds [Pt(en)$_2X$](ClO$_4$)$_2$ are all
single-peaked, \cite{W3143}
at $\hbar\omega\simeq 2.7\,\mbox{eV}$ for $X=\mbox{Cl}$ and
at $\hbar\omega\simeq 2.0\,\mbox{eV}$ for $X=\mbox{Br}$.
Figure \ref{F:OCOP} is reminiscent of these observations, whereas
Fig. \ref{F:OCIP} must be characteristic of the ladder system.
Although (bpym)[Pt(en)$X$]$_2$ and (bpy)[Pt(dien)Br]$_2$ are both shaped
like ladders, their electronic structures are distinct from each other,
as is shown in Fig. \ref{F:rho}.
In an IP-CDW state, every pair of Pt $d_{z^2}$ orbitals facing each other
across a ligand are well hybridized and split into their bonding
($d_\sigma$) and antibonding ($d_{\sigma^*}$) combinations with the help
of the bridging $\pi$ orbital.
The local density of states reveals a significant contribution of $\pi$
orbitals to the $d_{\sigma^*}$ bands.
The fully occupied $d_{z^2}(\mbox{Pt}^{2+})$ orbitals are much more
stabilized than the vacant $d_{z^2}(\mbox{Pt}^{4+})$ ones, that is to say,
$\varepsilon[d_{\sigma^*}(\mbox{Pt}^{4+},\bar{\mbox{Pt}}^{4+})]
-\varepsilon[d_{\sigma  }(\mbox{Pt}^{4+},\bar{\mbox{Pt}}^{4+})]
 \ll
 \varepsilon[d_{\sigma^*}(\mbox{Pt}^{2+},\bar{\mbox{Pt}}^{2+})]
-\varepsilon[d_{\sigma  }(\mbox{Pt}^{2+},\bar{\mbox{Pt}}^{2+})]$.
It is the broken electron-hole symmetry that unequalizes the optically
allowed excitations of two types.
Thus we find a double-peaked absorption band.
The essential $d$-$\pi$ hybridization is characteristic of
(bpym)[Pt(en)$X$]$_2$.
In an OP-CDW state, on the other hand, there hardly occurs interchain
hybridization of Pt $d_{z^2}$ orbitals and thus the main absorption band
of Pt character remains single-peaked.
The density of states is nothing more than a simple sum of poorly mixed
$d$, $p$, and $\pi$ orbitals.
(bpy)[Pt(dien)Br]$_2$ still has a strong resemblance to conventional
$M\!X$ single-chain compounds.

   $\sigma_\perp(\omega)$---With the $p$ and $\pi$ electrons included,
an absorption of Pt character in the rung direction is activated in an
IP-CDW state and is strengthened, roughly doubled, in an OP-CDW state.
However, it is still much less recognizable than that in the leg
direction.
Most of the spectral weight is distributed to the higher-energy region,
which is attributable to $\pi$-$d$ charge-transfer excitations.
The single-excitation CI scheme seems still incomplete but fully
demonstrates the crucial role of electronic correlations in reproducing
the observations quantitatively.
It may also be effective to take ligand $\pi^*$ orbitals into calculation.
Here we have discarded the vacant $\pi^*$ orbitals, on one hand assuming
them to be higher lying than Pt $d_{z^2}$ orbitals, and on the other hand
avoiding further increase of the number of parameters.
A pioneering density-functional study \cite{I063708} on
(bpym)[Pt(en)Cl]$_2$ proposes a level scheme of the bpym $\pi^*$ orbitals
being sandwiched between the $d_{z^2}(\mbox{Pt}^{2+})$ and
$d_{z^2}(\mbox{Pt}^{4+})$ bands.
Such a scenario looks consistent with our underestimation of
$\sigma_\perp(\omega)$ for (bpym)[Pt(en)$X$]$_2$ and may explain the
low-energy shoulder or foot of its widespread band.
On the other hand, the bpy $\pi^*$ orbitals are likely to lie above the
Pt $d_{z^2}$ bands, judging from Fig. \ref{F:OCOP}.

\section{Summary}

   (bpym)[Pt(en)$X$]$_2$ reveal themselves as novel
$d$-$p$-$\pi$-hybridized multiband ladder materials with a ground state of
the IP-CDW type, while (bpy)[Pt(dien)Br]$_2$ as a $d_{z^2}$-single-band
double-chain material with a ground state of the OP-CDW type, which is
reminiscent of conventional $M\!X$ chain compounds.
The two ground states are highly competitive and both materials sit in the
vicinity of the phase boundary.
An iodine derivative of the former compounds,
($\mu$-bpym)[Pt(en)I]$_2$I$_4\cdot 2$H$_2$O,
\cite{K7372} might have a ground state of OP-CDW character. \cite{K}

   There lie ahead fascinating topics such as quantum phase transitions
and nonlinear photoproducts in this geometrically designed Peierls-Hubbard
multiband system.
Palladium and nickel analogs as well as ligand substitution will
contribute toward realizing further density-wave states \cite{F044717}
possible in a multiorbital ladder lattice.
Photogenerated excitons and their relaxation channels were extensively
calculated for $M\!X$ \cite{M5758,M5763,S1605,I1088} and $M\!M\!X$
\cite{O250,O045122} chains and the predicted scenarios were indeed
demonstrated experimentally. \cite{O2023,T2169}
Photoexcited $M\!X$ ladders are more and more interesting.
Contrastive materials with IP-CDW and OP-CDW backgrounds have been
provided and identified.
The new stage is ready for further investigations.

\acknowledgments

   We are grateful to K. Iwano for fruitful discussion and valuable
comments on our calculation.
H. Matsuzaki and H. Okamoto have allowed and encouraged us to discuss
their elaborate optical observations.
Their kindness is greatly appreciated.
We further thank D. Kawakami, M. Yamashita, A. Kobayashi, and H. Kitagawa
for useful informations on their brandnew $M\!X$ ladder products.
This work was supported by the Ministry of Education, Culture, Sports,
Science and Technology of Japan.
\begin{figure}
\centering
\includegraphics[width=78mm]{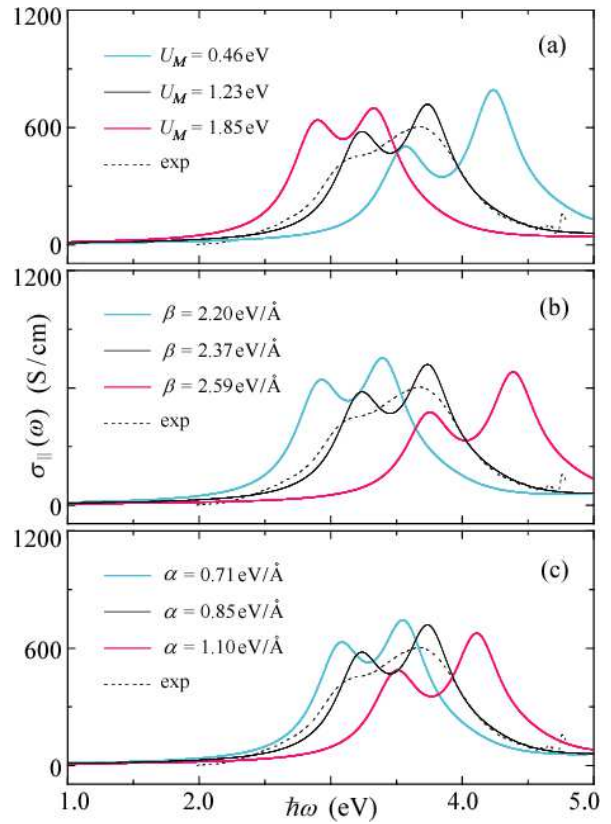}
\vspace*{-2mm}
\caption{(Color online)
         Single-excitation configuration-interaction calculations (solid
         lines) of the polarized optical conductivity spectra parallel to
         ladder legs for IP-CDW states in comparison with an experimental
         observation (a dotted line) of (bpym)[Pt(en)Cl]$_2$, where $U_M$
         (a), $\beta$ (b), and $\alpha$ (c) are tuned, while the rest are
         fixed at the set A in Table \ref{T:MP}.}
\label{F:Mtuning}
\end{figure}

\begin{figure}
\centering
\includegraphics[width=78mm]{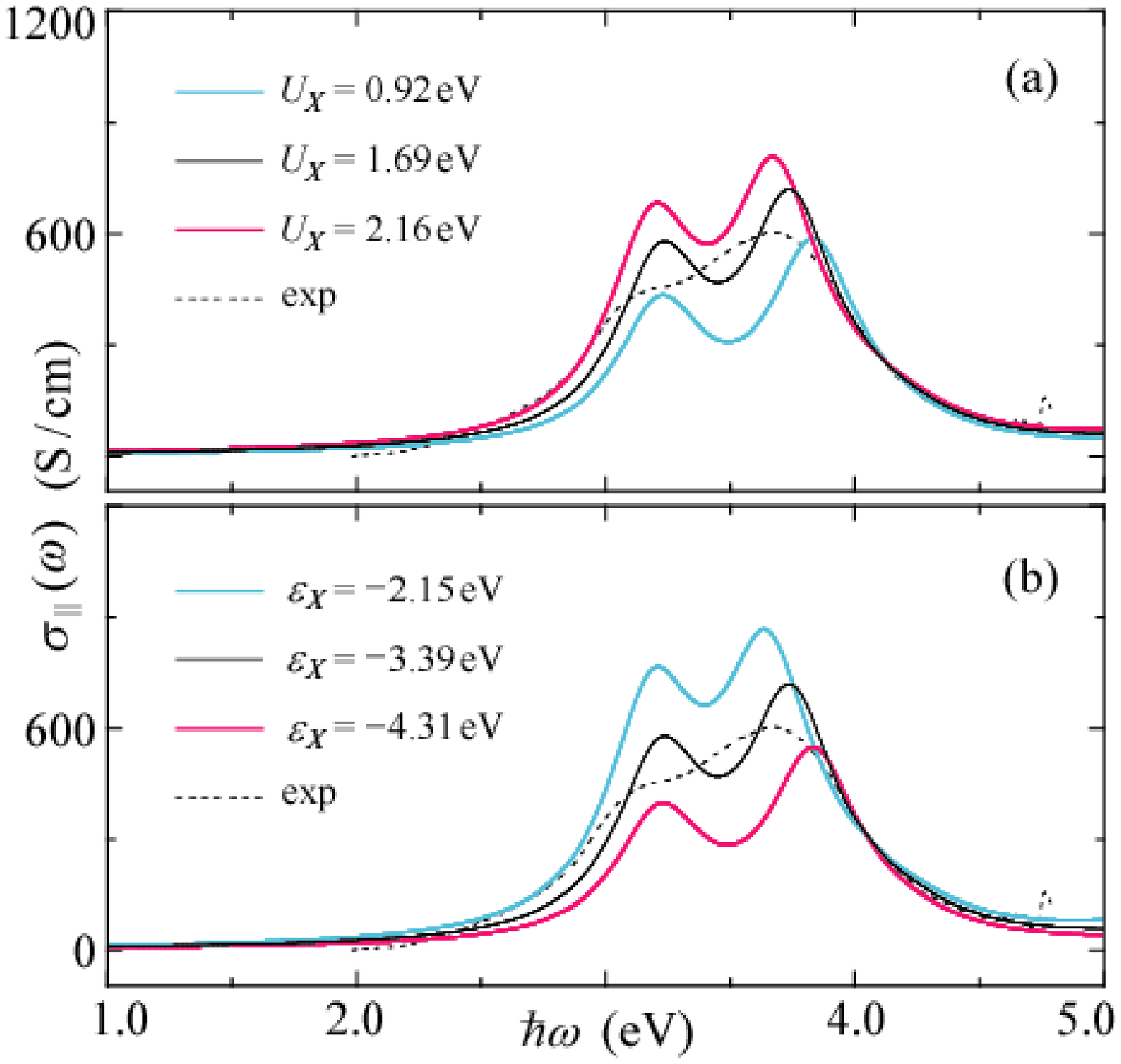}
\vspace*{-2mm}
\caption{(Color online)
         Single-excitation configuration-interaction calculations (solid
         lines) of the polarized optical conductivity spectra parallel to
         ladder legs for IP-CDW states in comparison with an experimental
         observation (a dotted line) of (bpym)[Pt(en)Cl]$_2$, where $U_X$
         (a) and $\varepsilon_X$ (b) are tuned, while the rest are fixed
         at the set A in Table \ref{T:MP}.}
\label{F:Xtuning}
\end{figure}

\begin{figure}
\centering
\includegraphics[width=78mm]{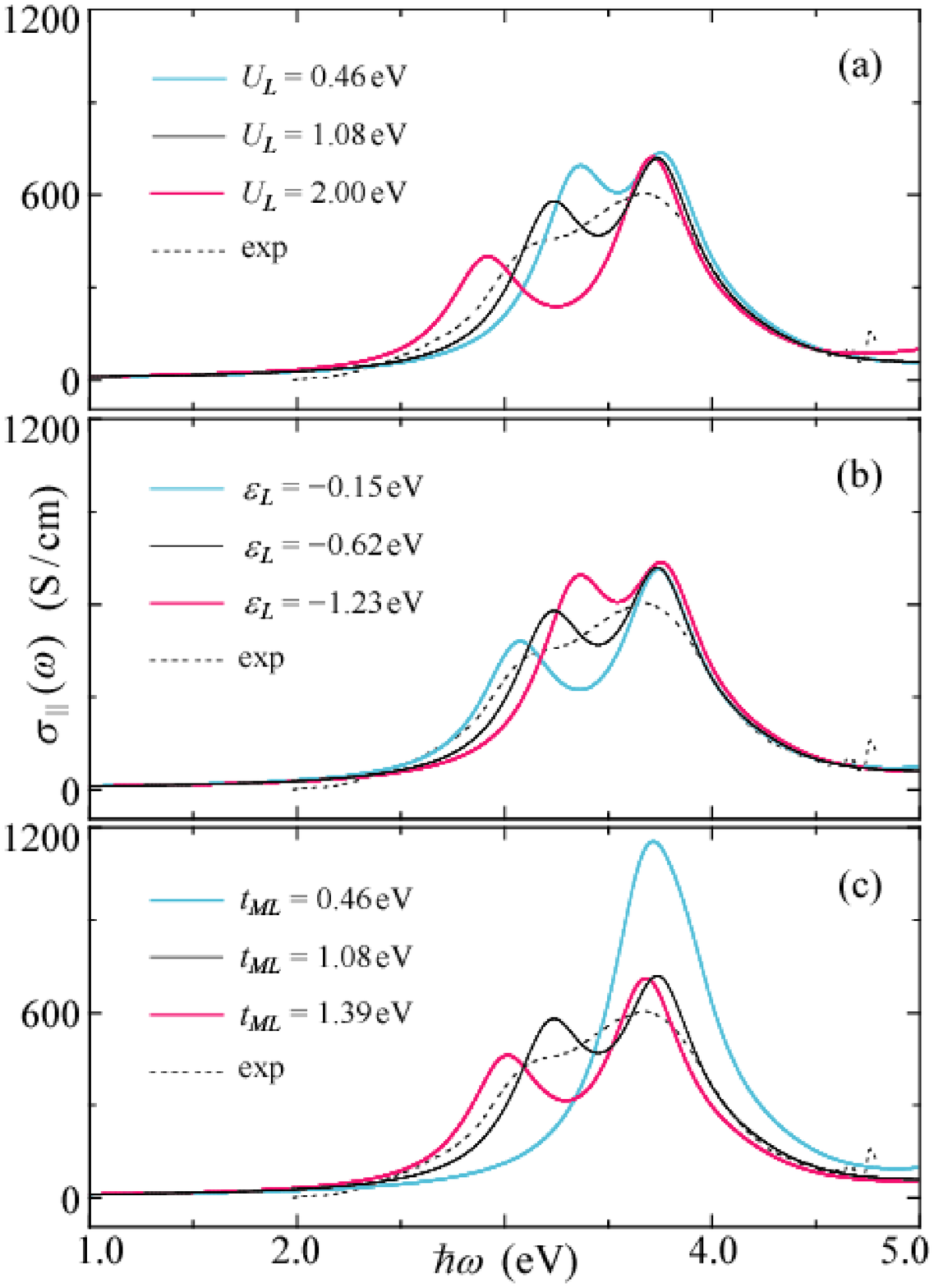}
\vspace*{-2mm}
\caption{(Color online)
         Single-excitation configuration-interaction calculations (solid
         lines) of the polarized optical conductivity spectra parallel to
         ladder legs for IP-CDW states in comparison with an experimental
         observation (a dotted line) of (bpym)[Pt(en)Cl]$_2$, where $U_L$
         (a), $\varepsilon_L$ (b), and $t_{M\!L}$ (c) are tuned, while the
         rest are fixed at the set A in Table \ref{T:MP}.}
\label{F:Ltuning}
\end{figure}

\begin{figure}
\centering
\includegraphics[width=78mm]{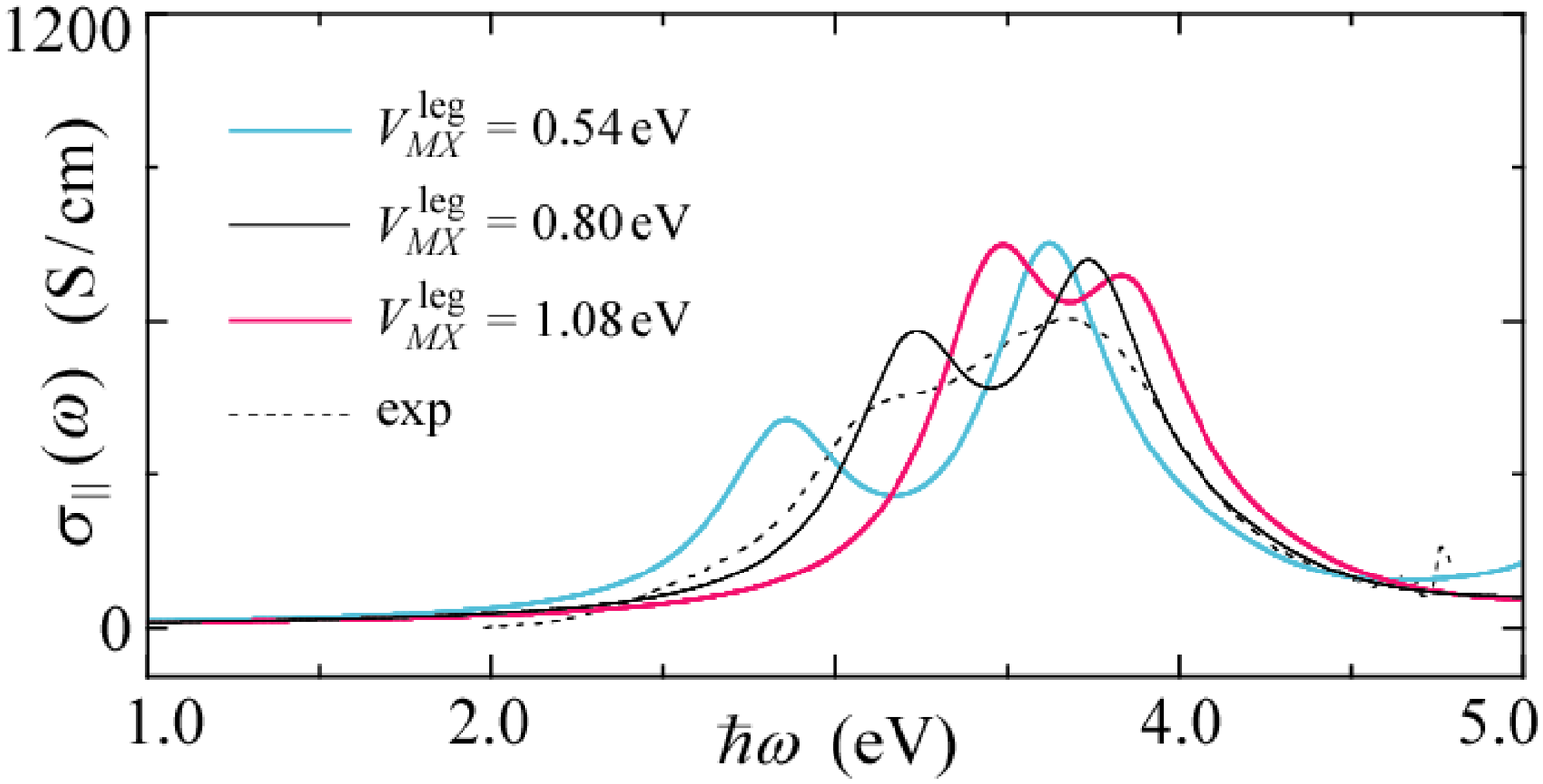}
\vspace*{-2mm}
\caption{(Color online)
         Single-excitation configuration-interaction calculations (solid
         lines) of the polarized optical conductivity spectra parallel to
         ladder legs for IP-CDW states in comparison with an experimental
         observation (a dotted line) of (bpym)[Pt(en)Cl]$_2$, where all
         the parameters but $V_{M\!X}^{\rm leg}$ are fixed at the set A in
         Table \ref{T:MP}.}
\label{F:Vtuning}
\end{figure}

\begin{appendix}
\section{On the Parameter Tuning}
\label{A:tuning}

   We have reached the best solutions in Figs. \ref{F:OCIP} and
\ref{F:OCOP} systematically tuning all the parameters.
The optical-conductivity spectral shape monotonically varies as we tune
each parameter.
We demonstrate the parameter tuning for (bpym)[Pt(en)Cl]$_2$ and discuss
what roles leading parameters play in reproducing the spectra.

   Varying $U_M$ slides, rather than deform, the spectrum
(Fig. \ref{F:Mtuning}).
With increasing $U_M$, the Peierls gap is reduced and any charge-transfer
excitation energy monotonically decreases.
The effect of electron-lattice interactions can be understood in the same
context.
The site-diagonal coupling constant $\beta$ straightforwardly stabilizes
a CDW on metal sites.
Considering $M$-$X$ charge-transfer energy gains, the site-off-diagonal
coupling constant $\alpha$ also stabilizes a site-diagonal CDW rather than
a site-off-diagonal (bond-centered) CDW, provided
$\varepsilon_M\neq\varepsilon_X$.
Both $\alpha$ and $\beta$ work against $U_M$.
All these parameters position the intrachain charge-transfer band.

   $U_X$ and $\varepsilon_X$ adjust the spectral weight of the main
absorption band originating from intrachain $M$-$X$ charge transfer
excitations (Fig. \ref{F:Xtuning}).
The oscillator strength of the charge-transfer band increases with
activated $p$ electrons.
Increasing $U_X$ induces oxidation of $X^-$ ions, while $\varepsilon_X$
approaching to $\varepsilon_M$ activates $d$-$p$ hybridization.
The spectral weight increases with increasing $U_X$ and decreasing
$\varepsilon_M-\varepsilon_X$.

   Parameters related to rung ligands control the structure of the main
absorption band (Fig. \ref{F:Ltuning}).
With increasing $U_L$, $\varepsilon_L$ approaching to $\varepsilon_M$,
and increasing $t_{M\!L}$, $d$-$\pi$ hybridization is encouraged.
Then the intrachain $d$ bands of Pt$^{2+}$ character split into their
bonding and antibonding combinations and the charge-transfer band is
doubly peaked.

   Finally we take a look at the effect of different-site Coulomb
interactions (Fig. \ref{F:Vtuning}).
The HF decomposition of any Coulomb term reminds us that the Coulomb
interaction originates from electron hopping between the relevant sites.
$V_{M\!X}^{\rm leg}$ indeed modulates the band gap in the same way as
$\alpha$ at the HF level.
However, the configuration interaction restructures the charge-transfer
band and drastically changes its double-peaked features.
With $V_{M\!X}^{\rm leg}$ large enough, the lower-energy absorption can
even be stronger than the higher-energy one.

   $V_{M\!M}^{\rm rung}$ and $V_{M\!M}^{\rm diag}$ are also important
Coulomb interactions, though they act on the next-nearest-neighbor sites.
They highly compete with each other for the ground-state valence
arrangement.
Therefore, these parameters are much less tunable and determined with
smaller uncertainty.
Thus and thus, we are led to the parametrization in Table \ref{T:MP} and
theoretical findings in Figs. \ref{F:OCIP} and \ref{F:OCOP}.
Considering the structural data as well, there is no better solution
within the present modeling.
\end{appendix}

\end{document}